# Energy Supply for Heating the Slow Solar Wind Observed by Parker Solar Probe between 0.17 and 0.7 au

Honghong Wu[1], Chuanyi Tu[1], Xin Wang[2], Jiansen He[1], and Liping Yang[3]
[1] School of Earth and Space Sciences, Peking University, Beijing, People's Republic of China; honghongwu@pku.edu.cn
[2] School of Space and Environment, Beihang University, Beijing, People's Republic of China
[3] SIGMA Weather Group, State Key Laboratory for Space Weather, National Space Science Center, Chinese Academy of Sciences, Beijing, People's Republic of China

## Abstract

Energy supply sources for the heating process in the slow solar wind remain unknown. The Parker Solar Probe (PSP) mission provides a good opportunity to study this issue. Recently, PSP observations have found that the slow solar wind experiences stronger heating inside 0.24 au. Here for the first time we measure in the slow solar wind the radial gradient of the low-frequency breaks on the magnetic trace power spectra and evaluate the associated energy supply rate. We find that the energy supply rate is consistent with the observed perpendicular heating rate calculated based on the gradient of the magnetic moment. Based on this finding, one could explain why the slow solar wind is strongly heated inside 0.25 au but expands nearly adiabatically outside 0.25 au. This finding supports the concept that the energy added from the energy-containing range is transferred by an energy cascade process to the dissipation range, and then dissipates to heat the slow solar wind. The related issues for further study are discussed.

*Unified Astronomy Thesaurus concepts:* Slow solar wind (1873); Interplanetary turbulence (830); Solar coronal heating (1989)

## 1. Introduction

Solar wind heating has long been an intriguing problem in space plasma physics. Given that the magnetic moment is conserved in the double adiabatic theory of plasma thermal dynamics (Chew et al. 1956), the radial variations of perpendicular temperature and total temperature follow $T_\perp \propto r^{-2}$ and $T \propto r^\alpha$, where $\alpha = -4/3$. Helios observations (Marsch et al. 1982, 1983; Schwenn 1983) give a chance to study the solar wind heating effects during its expansion between 0.3 and 1 au. The fast solar winds tend to be heated significantly, resulting in $\alpha = -0.8$ (Freeman 1988; Totten et al. 1995; Hellinger et al. 2011; Perrone et al. 2019). Bavassano et al. (1982a, 1982b) showed radial variations of high-speed wind observed by Helios with the solar wind streams coming from the same source at different times. These results were explained by Tu et al. (1984) and Tu (1988) with a WKB-like turbulence cascade model. However, the slow solar winds expand almost adiabatically with $\alpha = -1.3$ to $-1.2$ (Freeman 1988). Hellinger et al. (2013) studied the slow solar wind thermal energetics and found that the protons need to be heated. Whether or not the low-speed wind experiences any heating is an important unsolved problem.

A related critical issue is the existence of a turbulence energy cascade process in the slow solar wind. The fluctuations in the slow wind streams may be considered to be a fully developed turbulence (Dasso et al. 2005; Howes et al. 2008), in which the energy-containing eddies at the largest scale supply the energy to the cascade process in the inertial range, and finally the energy is dissipated to heat the wind at the small scales. So the energy supply rate should be equal to the heating rate, in the sense of dynamic equilibrium. This concept was applied to the fast wind turbulence observed by Helios (Bavassano et al. 1982a, 1982b; Tu et al. 1984; Tu 1988). Bruno & Carbone (2013) pointed out that Tu et al. (1984) presented for the first time the decreasing of the low-frequency breaks $f_c$ during the evolution, which shows the transfer of a part of the energy-containing range with $-1$ spectral index to the inertial range with $-3/2$ (or $-5/3$) spectral index (Tu & Marsch 1995). However, when applying this turbulence evolution concept to the slow solar wind, one faces the difficulty of identifying a frequency range with $-1$ spectral index, and hence an inability to identify the low-frequency breaks $f_c$. The slow wind streams observed by the Helios spacecraft are usually short in time interval, and no low-frequency breaks could be identified on the power spectrum (Marsch & Tu 1990). Bruno et al. (2019) pointed out that slow streams (observed to date) had not shown a clear low-frequency break. They found that, only in very rare cases in the Wind 12 yr observations, the low-frequency breaks $f_c$ could be identified on the magnetic power spectra. It remains a mystery whether or not there exist an energy-containing range with $-1$ spectral index and an energy cascade process in the slow solar wind.

Recently, the Parker Solar Probe (PSP; Fox et al. 2016) observations provided a critical opportunity to further investigate the solar wind heating. The magnetic moment of the slow wind from 0.17 to 0.24 au is observed to increase fast indicating strong heating, while it is almost constant outside 0.24 au showing nearly adiabatic expansion (Huang et al. 2020). These observations raise a question why the heating behavior of the slow solar wind are different inside and outside 0.24 au, respectively. This question has not yet been solved although several studies related with the wind heating and the turbulence evolution have been published. The energy transfer rate in the inertial range was evaluated with the third-order moments and compared with the wind heating rate (Bandyopadhyay et al. 2020). The stochastic heating rate was evaluated based on the velocity fluctuations observed at the scale of proton gyroradius and also compared with some heating results (Martinović 2020). The low-frequency breaks appear on the magnetic power spectra (Chen et al. 2020). None of these papers studying the slow solar wind successfully address the problem why the slow solar wind experiences strong heating inside 0.24 au but not outside.

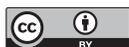

Here we present for the first time the radial gradient of the low-frequency breaks $f_c$ on the magnetic-trace power spectra in the slow solar wind. We propose a formula for calculating the energy supply rate associated with the decrease of $f_c$. We will show the consistency between the energy supply rate and the observed heating rate evaluated based on the radial gradient of the magnetic moment. This result supports the application of the general turbulence concept to the slow solar wind observed by PSP and explains the different heating rates inside and outside 0.24 au. This Letter is organized as follows. In Section 2, we describe PSP measurements. We provide equations to calculate the observed heating rate and derive the energy supply rate. We introduce a method to obtain the observed energy supply rate based on the radial dependence of the determined low-frequency break. In Section 3, we show the radial profiles of the temperature and heating rate. We also present the radial dependence of the low-frequency break and the energy supply rate. In Section 4, we discuss our results and draw our conclusions.

## 2. Data and Method

Here, we use the measurements from the first three orbits of PSP from 2018 October 31 to 2019 October 13. The magnetic field data are from the fluxgate magnetometer (MAG) in the FIELDS instrument suite (Bale et al. 2016). The plasma data are obtained from the PSP Solar Wind Electrons, Protons, and Alphas (SWEAP; Kasper et al. 2016) instrument suite. Proton velocity distribution function (VDF) moments are used, including proton density $n_p$, bulk velocity $V_{sw}$, and temperature $T_r$, as measured by the Solar Probe Cup (SPC; Case et al. 2020). The PSP mission design divides operations within an orbit into two phases: a solar encounter phase where the spacecraft is closer than 0.25 au, and a downlink/cruise phase outside the solar encounter region (Fox et al. 2016). The encounter phase operates at the highest sampling rates and the cruise phase operates at reduced sampling rates (Bale et al. 2016; Kasper et al. 2016).

The SPC measures the reduced VDFs and reports an effective thermal temperature that is a function of the orientation between the Cup's look direction and the ambient magnetic field, so that one can determine the parallel and perpendicular temperatures based on the varying orientation over short time intervals (Kasper et al. 2002, 2006). The effective thermal temperature is given by

$$T_r = T_\parallel (\boldsymbol{n} \cdot \boldsymbol{b})^2 + T_\perp [1 - (\boldsymbol{n} \cdot \boldsymbol{b})^2], \quad (1)$$

where $\boldsymbol{n}$ is the Cup's look direction, $\boldsymbol{b}$ is the magnetic field direction, $T_\parallel$ and $T_\perp$ are the parallel and perpendicular temperature, respectively. As SPC utilizes a single Sun-pointed look direction inside 0.7 au (Fox et al. 2016), $(\boldsymbol{n} \cdot \boldsymbol{b})^2 = (\boldsymbol{r} \cdot \boldsymbol{b})^2 = (B_r/B)^2$ ($B_r$ is the radial magnetic field and $B$ is the magnetic field magnitude). During the short intervals, the successive radial measurements report the effective thermal temperatures $T_r$ with different $(B_r/B)^2$. We perform least-square linear fits for $T_r$ and $(B_r/B)^2$ and obtain $T_\perp$. We utilize the criteria to obtain the temperature anisotropy as in Huang et al. (2020). Thus the time resolution of the fitted perpendicular temperature data is 10 s for Encounter phase and 1 minute for Cruise phase inside 0.7 au.

We cut the data into 6 minute intervals without overlapping and require that each interval contains more than four perpendicular temperature data points. We average $n_p$, $V_{sw}$, $T_\perp$ and $\boldsymbol{B}$ for each interval. The data in the fast solar wind cannot cover the whole 0.17–0.7 au region. We analyze 4131 intervals of the slow solar wind ($V_{sw} < 400$ km s$^{-1}$).

The observed perpendicular heating rate per unit volume can be calculated as

$$H(r) = n_p m_p \, Q_{\perp\,\mathrm{emp}} = n_p m_p \, BV_{sw} \frac{d}{dr}\left(\frac{\kappa T_\perp}{m_p B}\right), \quad (2)$$

where $\kappa$ is the Boltzmann constant, $m_p$ is the proton mass, and $Q_{\perp\,\mathrm{emp}}$ is the observed perpendicular solar wind heating rate per unit mass (Chandran et al. 2011; Bourouaine & Chandran 2013; Martinović et al. 2020). If we assume the radial dependence as $T_\perp = T_{\perp 0} \, r^{\alpha_T}$, $n_p = n_{p0} \, r^{\alpha_n}$, $B = B_0 \, r^{\alpha_B}$, we obtain the same formula as derived by Tu (1988):

$$H(r) = \frac{V_{sw} \, n_{p0} \, \kappa_B \, T_{\perp 0}}{1 \text{ au}} (\alpha_T - \alpha_B) \, r^{\alpha_n + \alpha_T - 1}. \quad (3)$$

The radial count densities variance of $T_\perp$ in Figure 1 (c) shows that the radial profile has two parts: 0.165 au $< r <$ 0.25 au and 0.25 au $< r <$ 0.7 au. We perform least-square linear fits for radial trends of $n_p$, $B$, $T_\perp$, and magnetic moments $T_\perp/B$ in the log–log space for each part. We obtain the fitted parameters ($y_0$ and $\alpha$) for the relationships $y = y_0 r^\alpha$, where $y$ represents respectively $n_p$, $B$, $T_\perp$, $T_\perp/B$. Using these fitted parameters and Equation (3), we calculate the radial heating rate profiles.

The wave spectral transfer equation has been evaluated by Tu et al. (1984) and its isotropic form is (Tu & Marsch 1995)

$$\frac{1}{A}\frac{\partial}{\partial r}\left(A(V_{sw} + V_{Ar})\frac{P(f,r)}{4\pi}\right) + \frac{P(f,r)}{8\pi}\frac{1}{A}\frac{\partial}{\partial r}(AV_{sw})$$
$$= -\frac{\partial}{\partial f}\frac{F(f,r)}{4\pi}, \quad (4)$$

where $A$ is the area function of the flow, $V_{Ar}$ is the radial Alfvén speed, $P(f, r)$ is the magnetic power spectrum as a function of frequency $f$ and heliocentric distance $r$, and $F(f, r)$ is the spectral flux function through the cascade process. Equation (4) describes the radial evolution of $P(f, r)$. Performing an integral over Equation (4) from the low-frequency break $f_c$ to the high-frequency break $f_b$, one may get

$$\frac{1}{A}\frac{\partial}{\partial r}\left[A(V_{sw} + V_{Ar})\int_{f_c}^{f_b}\frac{P(f,r)df}{4\pi}\right]$$
$$+ \int_{f_c}^{f_b}\frac{P(f,r)df}{8\pi}\frac{1}{A}\frac{\partial}{\partial r}(AV_{sw}) = \frac{F(f_c,r)}{4\pi} - \frac{F(f_b,r)}{4\pi}$$
$$+ (V_{sw} + V_{Ar})\frac{P(f_b,r)}{4\pi}\frac{df_b}{dr} - (V_{sw} + V_{Ar})$$
$$\times \frac{P(f_c,r)}{4\pi}\frac{df_c}{dr}. \quad (5)$$

The left side of Equation (5) is the WKB formula for Alfvén waves. The boundary terms at high-frequency break $f_b$ are $\frac{F(f_b, r)}{4\pi} - (V_{sw} + V_{Ar}) \frac{P(f_b, r)}{4\pi} \frac{df_b}{dr}$, which are the energy loss from the inertial range. The first term $\frac{F(f_b, r)}{4\pi}$ represents the energy flux cascading from the inertial range to the dissipation range, and the second term $(V_{sw} + V_{Ar}) \frac{P(f_b, r)}{4\pi} \frac{df_b}{dr}$ represents the energy transfer rate by the decreasing of $f_b$ when the wind expands. The first term dominates the second term and is consistent with the wind heating rate in the high-speed wind observed by Helios (Tu 1988). The boundary terms at $f_c$ are $\frac{F(f_c, r)}{4\pi} - (V_{sw} + V_{Ar}) \frac{P(f_c, r)}{4\pi} \frac{df_c}{dr}$, which are the energy supply into the inertial range. The energy flux $\frac{F(f_c, r)}{4\pi}$, which cascades into the inertial range from the low-frequency boundary, is also shown by Tu (1988) and is much lower than $\frac{F(f_b, r)}{4\pi}$. $F(f_c, r)/(4\pi)$ represents the energy flux cascading from the $-1$ frequency range into the inertial range. $F(f_c, r)/(4\pi)$ is difficult to be determined by the data analyses because it is hard to identify the real nonlinear process. However, at the frequency near $f_c$, the timescale is about $10^3$–$10^4$ s and the self-correlation coefficient of the magnetic fluctuations at this scale is low, indicating absent or weak nonlinear interactions. In addition, the spectrum is flat, indicating that the fluctuations may be uncorrelated and random. As a result, from general consideration we may expect that $F(f_c, r)/(4\pi)$ is very low or absent. The exact value of $F(f_c, r)/(4\pi)$ should be checked by future studies. At the present, the energy supply rate per unit volume may be evaluated as

$$S(r) = -(V_{sw} + V_{Ar}) \frac{P(f_c, r)}{4\pi} \frac{df_c}{dr}, \qquad (6)$$

where $\frac{df_c}{dr}$ is the radial gradient of $f_c$. The energy supply rate $S(r)$ describes the energy lost from the $-1$ spectrum and supplied into the inertial range during the radial evolution of the solar wind turbulence. This energy transfer process may be simply understood in the following way. We may neglect the width of the frequency band, in which the $-1$ range transits to the inertial range, and assume that the spectrum has two parts with spectral index $-1$ and $-3/2$ (or $-5/3$), respectively. With the radial evolution of the turbulence, the low-frequency break $f_c$ at the intersection of the two parts moves toward lower frequencies. During this process the high-frequency end of the $-1$ spectral range continues to define the low-frequency boundary of the $-3/2$ (or $-5/3$) spectral range. The new frequency band added to the higher frequency range may be marked as $df_c$. The energy contained in this frequency band is $2\frac{P(f_c,r)}{8\pi}df_c$. The factor of 2 is from consideration of the equipartition between kinetic and magnetic energy of Alfvénic waves. The time period $dt$, for $f_c$ to decrease to $(f_c - df_c)$, is $dr/(V_{sw} + V_{Ar})$. Then the formula $\frac{2P(f_c,r)}{8\pi}\frac{-df_c}{dt}$ is the energy transfer rate from the $-1$ spectral range to the the $-3/2$ (or $-5/3$) spectral range.

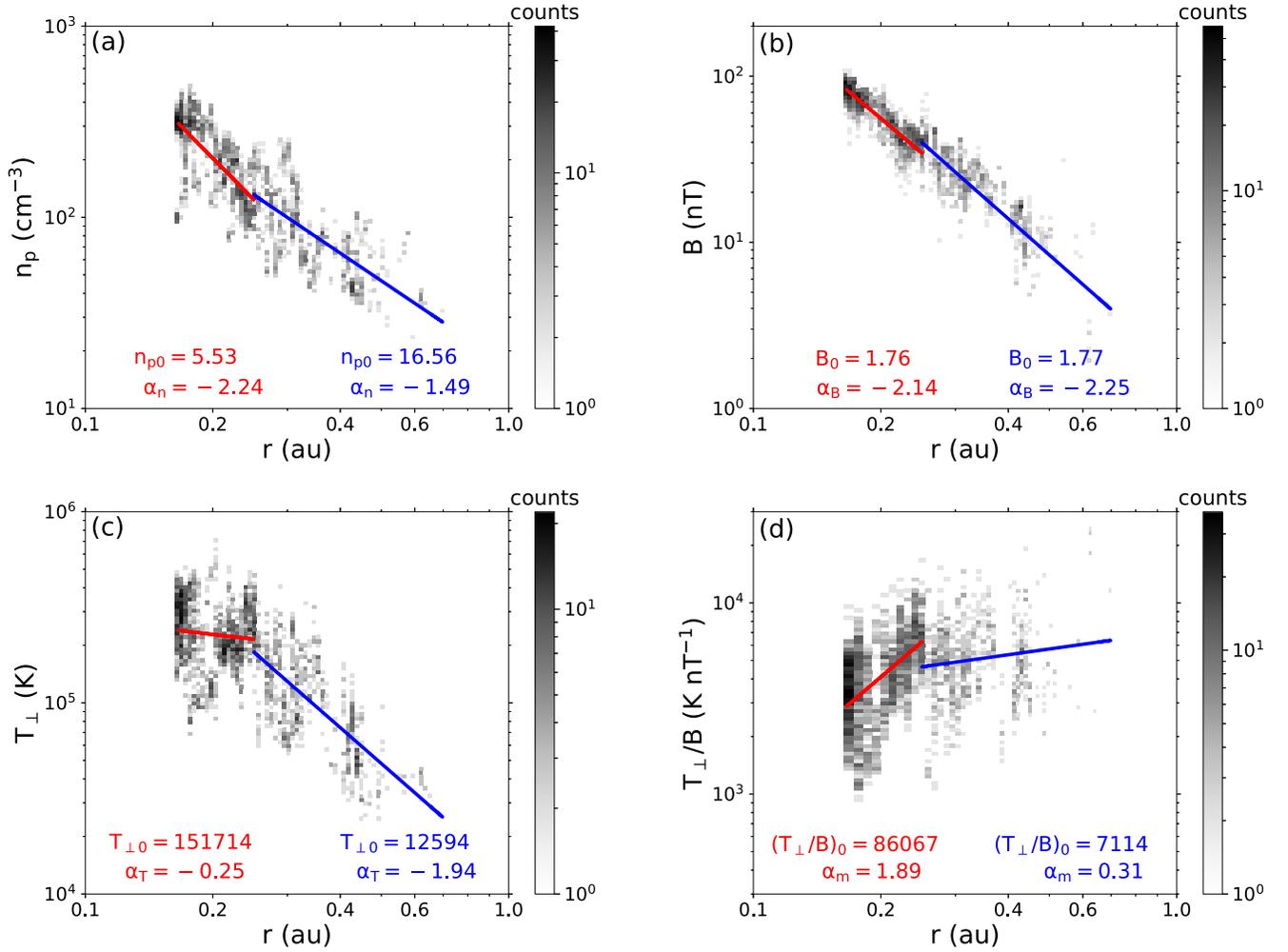

**Figure 1.** (a) Proton number density $n_p$, (b) magnetic field magnitude $B$, (c) perpendicular temperature $T_\perp$, and (d) magnetic moments $T_\perp/B$ of slow solar wind as a function of heliocentric distance. The gray color maps illustrate the observed count densities. The red and blue lines represent the least-square linear fitted relationships $y = y_0 r^\alpha$ with the fitted parameters ($y_0$ and $\alpha$) shown in two regions 0.165 au $< r <$ 0.25 au and 0.25 au $< r <$ 0.7 au, respectively. $y$ represents, respectively, the $n_p$, $B$, $T_\perp$, ($T_\perp/B$) in the four panels. There are 2410 counts inside 0.25 au and 1721 counts for 0.25 au $< r <$ 0.7 au.

We divide the MAG data into one-day intervals. We require more than 195,779 data points in each interval and the averaged bulk velocity smaller than 400 km s$^{-1}$. We linearly interpolate the data gaps and obtain the magnetic trace power spectrum PSD$_B$ by Fourier transform and smooth them by averaging over a sliding window of a factor of 2. We plot PSD$_B(f)$ separately in three groups according to $r$ where the spectra were observed, in the three panels of Figure 2. In each group, we can locate approximately the low-frequency break $f_c$ positions by eye, which are shown in Figure 2 by the red solid line on the power spectra. We calculate the radial gradient $\frac{df_c}{dr}$ using the values ($f_c$, $r$) corresponding to the two ends of the red solid line despite of the variations caused by the longitude differences. In this way, we obtain three gradients of low-frequency breaks for three radial distance regions: $-0.0075$ Hz au$^{-1}$ for 0.165 au $< r <$ 0.25 au, $-0.001$ Hz au$^{-1}$ for 0.25 au $< r <$ 0.45 au, and $-0.00012$ Hz au$^{-1}$ for 0.45 au $< r <$ 0.70 au.

We also obtain the corresponding intersections between the red solid line and every spectrum for each group. Thus we obtain 85 (24 + 24 + 37) pairs of ($f_c$, PSD$_B(f_c)$) with 85 heliocentric distances $r$. We calculate the energy supply rate using the corresponding 85 PSD$_B(f_c)$ values and the three radial gradient $\frac{df_c}{dr}$ in each region. Here we use $V_{sw} + V_{Ar} = 370$ km s$^{-1}$, which is the averaged value of 85 one-day intervals in the slow solar wind. The results are presented in the next section.

## 3. Results

Figure 1 shows the radial dependence of multiple variables for the slow solar wind. The color maps present the observed counts. The linear fits in log–log space are shown by the red and blue lines, respectively, with the corresponding fitted parameters. The radial dependences of $T_\perp$ and $T_\perp/B$ have two trends: inside 0.25 au, the temperature decreases much more slowly than the magnetic field magnitude $B$ and the magnetic moment $T_\perp/B$ increases with the increasing $r$; outside 0.25 au, the temperature decreases more quickly and $T_\perp/B$ increases more slowly than inside. These imply that the slow solar wind experiences strong heating inside 0.25 au and expands almost adiabatically outside 0.25 au. The continuously changes of $n_p$ and $B$ suggest that the observations represent mainly radial variation, but still contain some variations due to the azimuthal differences and fluctuations.

In Figure 2, we demonstrate the low-frequency breaks $f_c$ in the slow solar wind and the corresponding power spectra PSD$_B(f_c)$ in three different $r$ regions. It is clear that there are low-frequency breaks $f_c$ in those power spectra where the energy-containing range and the inertial range transit. $f_c$ moves to lower frequencies as the spectra evolve with the radial distance. This indicates that there are turbulence cascade processes in the slow solar wind as well. The power-law slopes gradually changes from $-3/2$ at 0.166 au to $-5/3$ at 0.7 au, as in Chen et al. (2020).

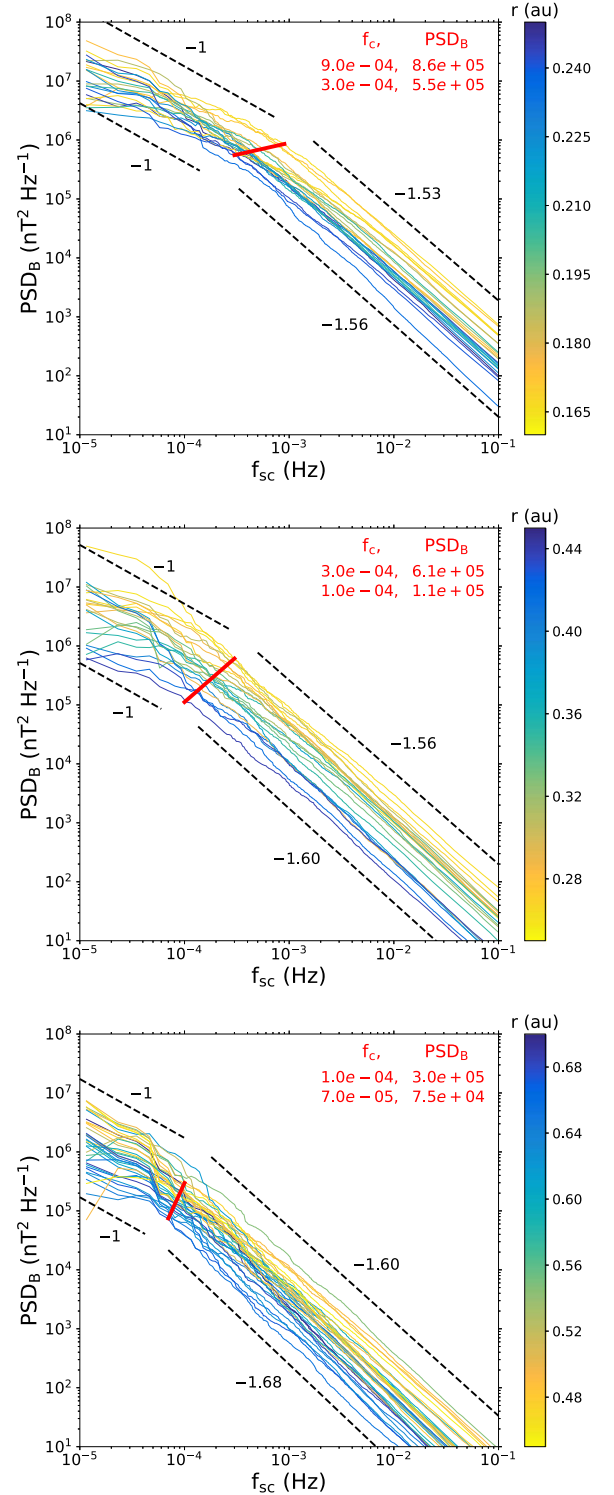

**Figure 2.** Magnetic trace power spectra (PSD$_B$) of the slow solar wind at different radial distances $r$. The top, the middle, the bottom panels represent 24, 24, and 37 spectra for 0.165 au $< r <$ 0.25 au, 0.25 au $< r <$ 0.45 au, and 0.45 au $< r <$ 0.7 au, respectively. The black dashed lines with marked power-law slopes illustrate the two spectral frequency ranges: the energy-containing range and the inertial range. The red solid line shows approximately the positions of the low-frequency break points on the spectra. The values of the frequency $f_c$ and the PSD$_B$ corresponding to the two ends of the red solid line are shown in red color digitals on the top-right corner in each panel.

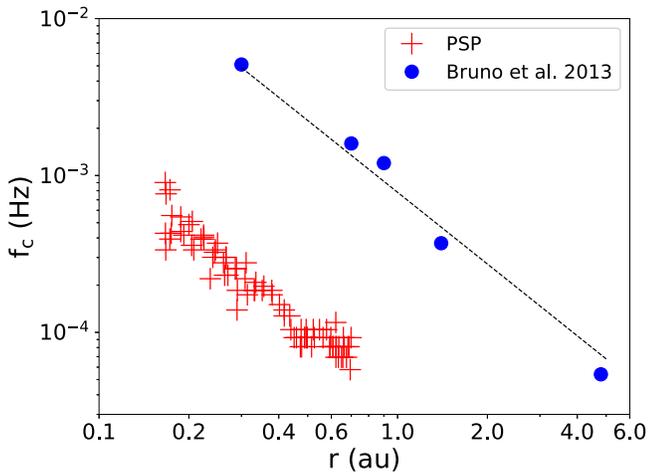

**Figure 3.** Radial variations of the low-frequency breaks in the slow solar wind (red plus symbols) obtained by PSP observations. The blue dots are adapted for comparison from Bruno & Carbone (2013) and represent the low-frequency breaks in the fast solar wind.

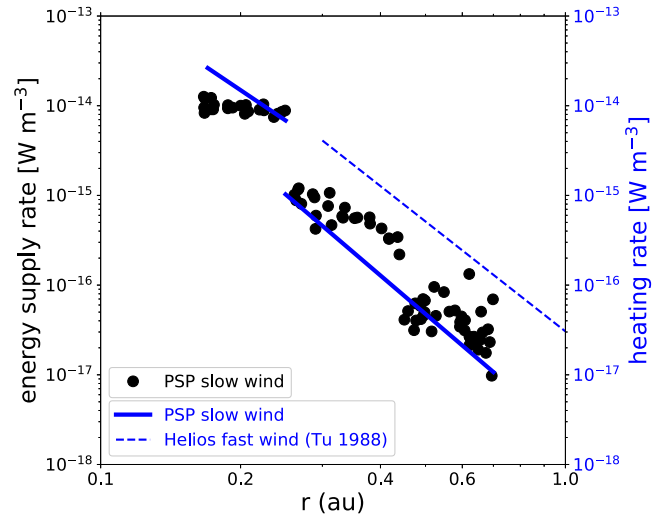

**Figure 4.** Dependences on radial distance of energy supply rate (black dots) and perpendicular heating rate (blue solid lines; $\propto r^{-3.49}$ inside 0.25 au and $\propto r^{-4.43}$ outside 0.25 au) calculated from PSP observations for slow solar wind. The blue dashed line indicates the perpendicular heating rate ($\propto r^{-4.07}$) obtained by Helios measurements for fast solar wind adapted from Tu (1988).

We present the 85 low-frequency breaks $f_c$ at 85 heliocentric distance $r$, respectively, in the slow solar wind shown by the red plus symbols in Figure 3. The blue dots are $f_c$ in the fast solar wind adapted from Bruno & Carbone (2013) for comparison. The low-frequency breaks in the slow solar wind is about an order of lower than that in the fast solar wind at the same radial distance. The gradient of $f_c$ presents a clear evolution: $f_c$ changes faster inside 0.25 au than outside, which may results in a stronger energy supply rate.

We show the heating rate $H(r)$ (blue) and energy supply rate $S(r)$ (black) in Figure 4. Inside 0.25 au, we have $H(r) \propto r^{-3.49}$ and, outside 0.25 au, we have $H(r) \propto r^{-4.43}$. The heating rate profile of the slow solar wind inside 0.25 au is comparable with the expected extension of the fast solar wind profile based on the Helios observation outside (see the blue solid line inside 0.25 au and the blue dashed line). The heating rate inside 0.25 au is considerably higher than that beyond 0.25 au. When the energy supply rate is high, the slow wind is heated strongly. When the energy supply rate is extremely low outside 0.25 au, the slow wind expands adiabatically. The energy supply rate consistently matches the observed heating rate, which suggests that the solar wind heating energy is supplied by the energy lost from the range of $-1$ spectrum through an energy cascade process. We also notice that $H(r)$ is above $S(r)$ inside 0.25 au. This could be a result of the divergence of the heat flux, which may also play an important role in the increase of magnetic moment.

### 4. Discussion and Conclusions

We present for the first time in slow solar wind the radial profile of the low-frequency break $f_c$ on the magnetic power spectra, and show the first result for the radial evolution of energy supply rate for the slow solar wind. We provide a formula to evaluate the energy supply rate from the $-1$ frequency range to the inertial range, which relates to the decreasing of $f_c$. We find the comparability between the energy supply rate and the perpendicular heating rate calculated by the observed gradient of the magnetic moment. Inside 0.25 au, the value of $f_c$ is higher and its evolution is much faster, resulting in a much larger energy supply rate and stronger heating than outside 0.25 au. The different heating rates observed inside and outside 0.25 au by PSP (Huang et al. 2020) may be explained by the different energy supply rates.

These results are obtained directly from the observational data. The phenomenological findings may be understood using the concept of turbulence. The energy supply rate describes the rate of the energy transferred from the energy-containing range whose spectrum has $-1$ slope to the inertial range whose spectrum has $-3/2$ (or $-5/3$) slope. The energy is further transferred to the dissipation range through an energy cascade process, and finally heats the solar wind protons. The consistency between the energy supply rate and the perpendicular heating rate indicates that the latter is ultimately controlled by the decrease rate of $f_c$. This idea is inaccurate and unproven, and the existence of $f_c$ and its displacement is not necessary to have a cascade. A cascade process may still exist without the $-1$ frequency range. However, this idea is a possible understanding for the relationship between the displacement of $f_c$ and the increase of magnetic moments. The process is quite different from the decay of the energy-containing eddies as conceived in hydrodynamic theory. Bruno & Carbone (2013) pointed out when commenting on the model created by Tu et al. (1984), Tu (1988), and Tu & Marsch (1995) that "this low-frequency range is not separate from the inertial range, but becomes part of it as the turbulence ages. These observations cast some doubt on the applicability of the hydrodynamic turbulence paradigm to interplanetary magneto-hydrodynamic turbulence." They also pointed out that in hydrodynamic turbulence, the energy-containing eddies decay during the turbulent evolution but will not become part of the inertial range. Our results show that this special energy transfer of solar wind turbulence does not only exist in the fast wind, but also appears in the slow solar wind. This possibility needs to be explored in future theoretical and observational studies.

The −1 frequency break has been associated in many works to the Alfvénic properties of the fast wind. Matteini et al. (2018) found that the presence of the −1 frequency range could be a result of the saturation of the amplitude of the magnetic fluctuations in the fast wind. D'Amicis et al. (2019) reported that the slow solar wind with high Alfvénicity share common characteristics with the fast wind, and Perrone et al. (2020) further observed that Alfvénic slow wind shows a break between the inertial range and large scales where the magnetic fluctuations saturate. It seems plausible that this −1 frequency range is related to the observations of Alfvénic switchbacks (de Wit et al. 2020; Horbury et al. 2020; Mozer et al. 2020; McManus et al. 2020), close the Sun, that provide most of the energy-containing range in the slow wind. Later on other processes such as stream interactions and reconnection could furnish the large scales and the cascade (even at lower rates). However, the relationship between the −1 frequency break and Alfvénic properties is beyond the scope of this work and requires further study.

Our results, however, have some uncertainties and require further refinements. The PSP observations are not strictly in a radial stream, but at different radial distances and different longitudes, and hence in different flow streams. Because the magnetic power spectra in different flows are different even at the same radial distance, one may not easily obtain the exact radial evolution trend from PSP data. Nonetheless, we identify the radial trend of $f_c$ by eye separately in the three radial groups of spectra. In this way, we obtain the first estimate of radial gradients of $f_c$ in the slow solar wind, which have a power-law shape with respect to the heliocentric distance, similar to the result by Helios observations in the fast solar wind. This estimate is relatively rough, and more accurate statistical studies are needed. More data from future PSP orbits and Solar Orbiter is required (Müller et al. 2013).

We acknowledge the NASA Parker Solar Probe mission team and the SWEAP team led by J.C. Kasper, and the FIELDS team led by S.D. Bale, for the use of PSP data. The data used in this Letter can be downloaded from spdf.gsfc.nasa.gov. This work is supported by the National Natural Science Foundation of China under contract Nos. 41974198, 41674171, 41874199, 41574168, 41874200, 41774183, 41774157, 41974171, and 41861134033.

### ORCID iDs

Honghong Wu https://orcid.org/0000-0003-0424-9228
Chuanyi Tu https://orcid.org/0000-0002-9571-6911
Xin Wang https://orcid.org/0000-0002-2444-1332
Jiansen He https://orcid.org/0000-0001-8179-417X
Liping Yang https://orcid.org/0000-0003-4716-2958